%% file: main.tex
\documentclass[sigconf]{acmart}

\usepackage{booktabs} 

\copyrightyear{2019} 
\acmYear{2019} 
\setcopyright{acmcopyright}
\acmConference[ASPDAC '19]{ASPDAC '19: 24th Asia and South Pacific Design Automation Conference}{January 21--24, 2019}{Tokyo, Japan}
\acmBooktitle{ASPDAC '19: 24th Asia and South Pacific Design Automation Conference (ASPDAC '19), January 21--24, 2019, Tokyo, Japan}
\acmPrice{15.00}
\acmDOI{10.1145/3287624.3287662}
\acmISBN{978-1-4503-6007-4/19/01}

\usepackage[loading]{tracefnt}
\usepackage{algorithmicx}
\usepackage{algorithm}
\usepackage{algpseudocode}
\usepackage{multirow}
\usepackage{subcaption}
\usepackage{setspace}
\usepackage[utf8]{inputenc}
\usepackage{amsmath}
\usepackage{amsfonts}
\usepackage{amssymb}
\usepackage{graphicx}
\usepackage{color}
\usepackage{xcolor}
\usepackage{todonotes}
\usepackage{bm}
\usepackage{color}
\newcommand{\RN}[1]{%
  \textup{\uppercase\expandafter{\romannumeral#1}}%
}
\begin{document}
%
\title{Spectral Approach to Verifying Non-linear Arithmetic Circuits}

\author{Cunxi Yu, Tiankai Su, Atif Yasin, Maciej Ciesielski}
\affiliation{
  \institution{University of Massachusetts, Amherst}}
\email{ycunxi, ayasin, tiankaisu, ciesiel@umass.edu}

\renewcommand{\shortauthors}{C. Yu et al.}

\begin{abstract}
This paper presents a fast and effective computer algebraic method for analyzing and verifying non-linear integer arithmetic circuits using a novel algebraic spectral model. It introduces a concept of algebraic spectrum, a numerical form of polynomial expression; it uses the distribution of coefficients of the monomials to determine the type of arithmetic function under verification. In contrast to previous works, the proof of functional correctness is achieved by computing an algebraic spectrum combined with local rewriting of word-level polynomials. The speedup is achieved by propagating coefficients through the circuit using And-Inverter Graph (AIG) datastructure.
The effectiveness of the method is demonstrated with experiments including standard and Booth multipliers, and other synthesized non-linear arithmetic circuits up to 1024 bits containing over 12 million gates.
\end{abstract}

\maketitle




%

\input{intro}
\input{background}

\input{spectral-method}

\input{function-extraction}
\input{adder-tree-extract}

\input{results}
\input{conclusions}

%
%
%
%



  
\vspace*{-2mm}
\section*{Acknowledgment}
This work was supported by an award from the National Science Foundation, No. CCF-1617708. The authors thank Alan Mishchenko for his help in integrating the tool with ABC.
%

\bibliographystyle{ACM-Reference-Format}
\bibliography{verification_ycunxi,synthesis}
\end{document}

%% file: intro.tex
\section{introduction}

Importance of arithmetic verification problem grows with an increased use of arithmetic modules in modern systems, such as signal processing, security engineering, and cryptographic applications. 
There has been a considerable progress in formal verification of arithmetic designs in the last decade. In particular, computer algebra techniques that use polynomial representation of a gate-level arithmetic circuit, show significant advantages in analyzing arithmetic circuits \cite{STABLE:date11}\cite{ciesielski2015verification}\cite{yu:2016-tcad-verification}\cite{pruss2016TCAD:efficient}\cite{sayedformal:date-2016}\cite{ritirc2017column}. 
This is in contrast to other formal methods, such as BDDs or SAT, that rely on a strictly Boolean circuit representation. The verification problem using computer algebraic methods is typically formulated as a proof that the implementation satisfies the specification, which is solved by polynomial division or by algebraic rewriting.

The techniques that play a major role in synthesis and verification, are {\it abstraction} and {\it reverse engineering} \cite{soeken:2015simulationgraph}\cite{yu:2016-abstraction}. Formal verification techniques can benefit greatly from abstracting functionality of the circuits being verified. For example, word-level abstraction specifically focuses on extracting a word-level representation of the function implemented by a gate-level design, which can significantly reduce the complexity of verifying a large system. In the past, the verification and abstraction problems relied entirely on {\it explicit} functional methods. In this paper, we describe an {\it implicit} approach to verification and word-level {\it abstraction} of arithmetic circuits by introducing a novel representation, called \textit{algebraic spectrum}. 
We describe an efficient algorithm for constructing such a spectrum, a compact way to represent the polynomial model of the circuit.

%% file: background.tex
\section{Background} \label{sec:related-work}
 
\subsection{Formal Verification of Arithmetic Circuits}
Verification of arithmetic circuits is performed using a variant of combinational equivalence checking, referred to as \textit{arithmetic combinational equivalence checking} (ACEC) \cite{sayedformal:date-2016}. Several approaches have been applied to verify an arithmetic circuit against its functional specification, including decision diagrams, satisfiability, theorem proving, and computer algebra. Different variants of canonical, graph-based representations have been proposed, including Binary Decision Diagrams (BDDs) \cite{bryant:1986-bdd}, Binary Moment Diagrams (BMDs) \cite{bryant:tr97}, Taylor Expansion Diagrams (TED) \cite{ted:tcomp06}, and other hybrid diagrams.
While BDDs have been used extensively in logic synthesis, their applicability to verification of arithmetic circuits is limited by the prohibitively high memory requirements imposed by complex arithmetic circuits, such as multipliers. 
Boolean satisfiability (SAT) and satisfiability modulo theories (SMT) solvers have also been applied to solve ACEC problems \cite{goldberg2001using}. Several state-of-the-art SAT and SMT solvers have been applied to those problems, including MiniSAT\cite{sorensson:2005-minisat}, Lingeling\cite{biere2013lingeling}, Boolector \cite{niemetz:2015boolector}, and others. However, the complexity of ACEC for large arithmetic circuits has been shown to be extremely high \cite{pruss2016TCAD:efficient} \cite{yu:2016-tcad-verification}. Alternatively, the problem can be modeled as equivalence checking against an arithmetic specification given by a bit-vector formula, but the complexity of this method is the same as the ACEC method \cite{yu:2016-tcad-verification}. 

\vspace{-1mm}
\subsection{Computer Algebra Approach}

Computer algebra methods are considered to be best suited to solve arithmetic verification problems \cite{yu:2016-tcad-verification}\cite{ritirc2017column}. Using these methods, the verification problem is formulated as a proof obligation, stating that the implementation satisfies the specification \cite{STABLE:date11,ciesielski2015verification,yu:2016-tcad-verification,pruss2016TCAD:efficient,yu-aspdac-17,sayedformal:date-2016,,ritirc2017column}. Computer algebra offers a way to accomplish this using the theory of \textit{Gr{\" o}bner} basis and the \textit{ideal membership testing} to check if the specification belongs to the ideal generated by the implementation.
It can be solved by performing a series of divisions of the specification polynomial by a set of polynomials (bases) representing circuit components and checking if the remainder of the division reduces to zero. 

An alternative approach to arithmetic verification of gate-level circuits has been proposed using an algebraic rewriting technique, described in more detailes in the next section. With this approach, the polynomial representing the encoding of the primary outputs (the {\it output signature}) is transformed into a polynomial expressed in terms of the primary inputs (the {\it input signature}) \cite{ciesielski2015verification}. This method, in fact, extracts an arithmetic function implemented by the circuit, hence it is termed \textit{function extraction}. It has been successfully applied to standard, non-optimized 512-bit multipliers due to the simplification of polynomials achieved during rewriting \cite{yu:2016-tcad-verification} \cite{sayedformal:date-2016}. Although these approaches show good performance in verifying arithmetic circuits with well-defined structure, they suffer from polynomial size explosion when applied to synthesized and heavily bit-optimized gate-level netlists.

A comprehensive review of the state-of-the-art computer algebra methods for arithmetic circuit verification can be found in \cite{ritirc2017column}. The authors formally prove soundness and completeness of the two complementary approaches: the polynomial rewriting method of \cite{yu:2016-tcad-verification}\cite{sayedformal:date-2016} and the ideal membership testing of \cite{STABLE:date11}. The difficulties of verifying bit-optimized and technology mapped multipliers have been discussed as well. They also propose an incremental approach to arithmetic circuit verification by column-based polynomial reduction. In addition, computer algebra methods have been applied to logic debugging \cite{samaneh:2015-debug,farimah:2016-date, mahzoon2018combining,su2018computer} and approximations of arithmetic circuits \cite{su2018computer,froehlich2018approximate}.

\vspace{-2mm}
\subsection{Function Extraction using Algebraic Rewriting}\label{sec:extraction}
This section briefly reviews the function extraction technique that motivation our approach. It computes a unique bit-level polynomial function implemented by the circuit directly from its gate-level implementation \cite{ciesielski2015verification}. 
It uses an algebraic model of the circuit, with logic gates represented by the following algebraic expressions, with circuit signals treated as {\it Boolean} variables.
\begin{equation} 
\scriptsize
\begin{split}
\neg a = 1 - a \\
~~a \wedge b = a\cdot b \\
a \vee b = a + b - a\cdot b \\
a \oplus b = a + b -2 a \cdot b
\end{split}
\label{eq:Boolean-poly}
\end{equation}

Functional correctness of the circuit is proved by rewriting the word-level expression of the output signature, $Sig_{out}$, into a word-level expression at the primary inputs (PI), the input signature, $Sig_{in}$. The rewriting process successively applies Eq. (\ref{eq:Boolean-poly}), combined with algebraic simplification of the polynomial, to arrive at each step at a unique polynomial expression. Specifically, such an expression is a pseudo-Boolean polynomial in the variables associated with the set of signals separating primary inputs from primary outputs (PO), referred to as a {\it cut}. 
The rewriting is performed in reverse-topological order, from PO to PI: once a given variable (output of a gate) is substituted by an algebraic expression of the gate inputs, it will be eliminated from the current cut expression and will never appear again. As a result, the final polynomial ($Sig_{in}$) is expressed only in the primary input variables, and hence provides the function computed by the circuit.

This paper describes a novel and more efficient approach to function extraction by applying two new concepts: 1) {\it generating polynomial coefficients} without explicit polynomial rewriting, using AIG traversal; and 2) {\it spectral analysis} to reason about the function of the intermediate polynomials by analyzing their coefficients.

%% file: spectral-method.tex
\vspace{-3mm}
\section{Spectral Method} \label{sec:spectral-method}
%
%
\subsection{Algebraic Spectrum} \label{sec:spectrum}
\begin{figure}[t]
  \centering
    \includegraphics[width=0.45\textwidth]{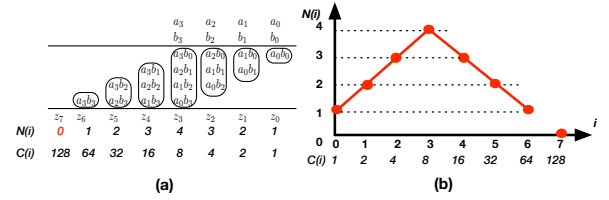} \\ 
     \vspace*{-2mm}
   \caption{Spectrum of a four-bit Multiplier}
    \label{fig:4bitMult}
  \end{figure}

Consider the $n$-bit integer multiplication scheme, shown in Figure \ref{fig:4bitMult}(a) for $n=4$. The ovals represent partial product terms that are added column-wise for each bit of the result. Let $i$ be a bit position of the result, $i = 0, ..., 2n-1$.
Note that $N(2n-1)$=0 since there are no partial product with coefficient of $2^{2n-1}$. 

Let $C_i=2^i$ be the coefficient associated with column $i$ of the result, and let $N_i$ be the number of product terms added at that bit position. The polynomial expression corresponding to the encoded word-level result is then:

\begin{equation} 
\scriptsize
F = \sum_{j=0}^{n-1} 2^j a_j \cdot \sum _{k=0} ^{n-1} 2^k b_k = \sum_{j=0}^{n-1}\sum_{k=0}^{n-1}2^{j+k} (a_j b_k)
\label{eq:spec-multiplication}
\end{equation} 
It is easy to see that each monomial $2^{j+k}a_jb_k$, for any pair of values of $j, k$, 
has the same coefficient, $C_i=2^{j+k}$, where $i=j+k$. The number of monomials with coefficient $C_i$ are represented using $N_i$.
For example, for a 2-bit unsigned multiplier with output $F=(a_0+2a_1)(b_0+2b_1) = a_0b_0 +2a_0b_1 +2a_1b_0 +4a_1b_1$, 
there is one monomial with coefficient $2^0$=1, two monomials with coefficient $2^1$=2, and one monomial with coefficient $2^2$=4. Hence, the set of coefficients for this polynomial, listed in the {\it increasing order} of coefficient value,  is $C=\{1, 2, 2, 4 \}$ and the set $N=\{ 1, 2, 1\}$.

Similarly, for the 4-bit multiplier shown in Figure \ref{fig:4bitMult}, 
we have: $N = \{1,2,3,4,3,2,1\}$, where the values of $N_i$ are listed in the increasing order of the output bits, from LSB to MSB. 
%
In general, the value of $N_i$ for an $n$-bit multiplier, with bits $i=0,...,2n-2$, can be computed as follows:
\begin{equation} 
\small
N_i =
  \begin{cases}
    i+1 & \quad \text{if } ~i \text{$\leq ~n - 1$}\\
    2n-1-i  & \quad \text{if } ~i \text{$\geq ~n$}\\
  \end{cases}
\label{eq:mult-spectrum-formula}
\end{equation}
%
The distribution of coefficients values $N_i(C_i)$ defines the algebraic spectrum of the polynomial and can be used to determine the type of the arithmetic function under investigation.

{\bf Definition 1:}
Given a polynomial $P = \sum C_i p_i$, where $C_i$ is an integer coefficient and $p_i$ is a monomial, product of some variables. Let $C=\{C_i\}$ be the set of coefficients of $P$ and let $N_i$ represent the number of product terms $p_i$ with the same coefficient $C_i$.
The {\it algebraic spectrum} $S$ for polynomial $P$ is then defined as an ordered set of pairs $(N_i,C_i)$, for all {\it distinct} values of coefficients $C_i$. That is, $S = \{(N_i,C_i)\}$. 
{\bf Example 1}: 
Let  $P = 3p_3 + 4p_2 + 4p_4 + 6p_1$, with monomials ordered by increasing values of its coefficients, 
Then the set of distinct coefficients is $C= \{3, 4, 6\}$ and the spectrum $S = \{ (1,3), (2,4), (1,6) \}$.

\begin{figure}[h]    
    \centering
    \begin{subfigure}[t]{0.23\textwidth}
        \centering
        \includegraphics[width=\textwidth]{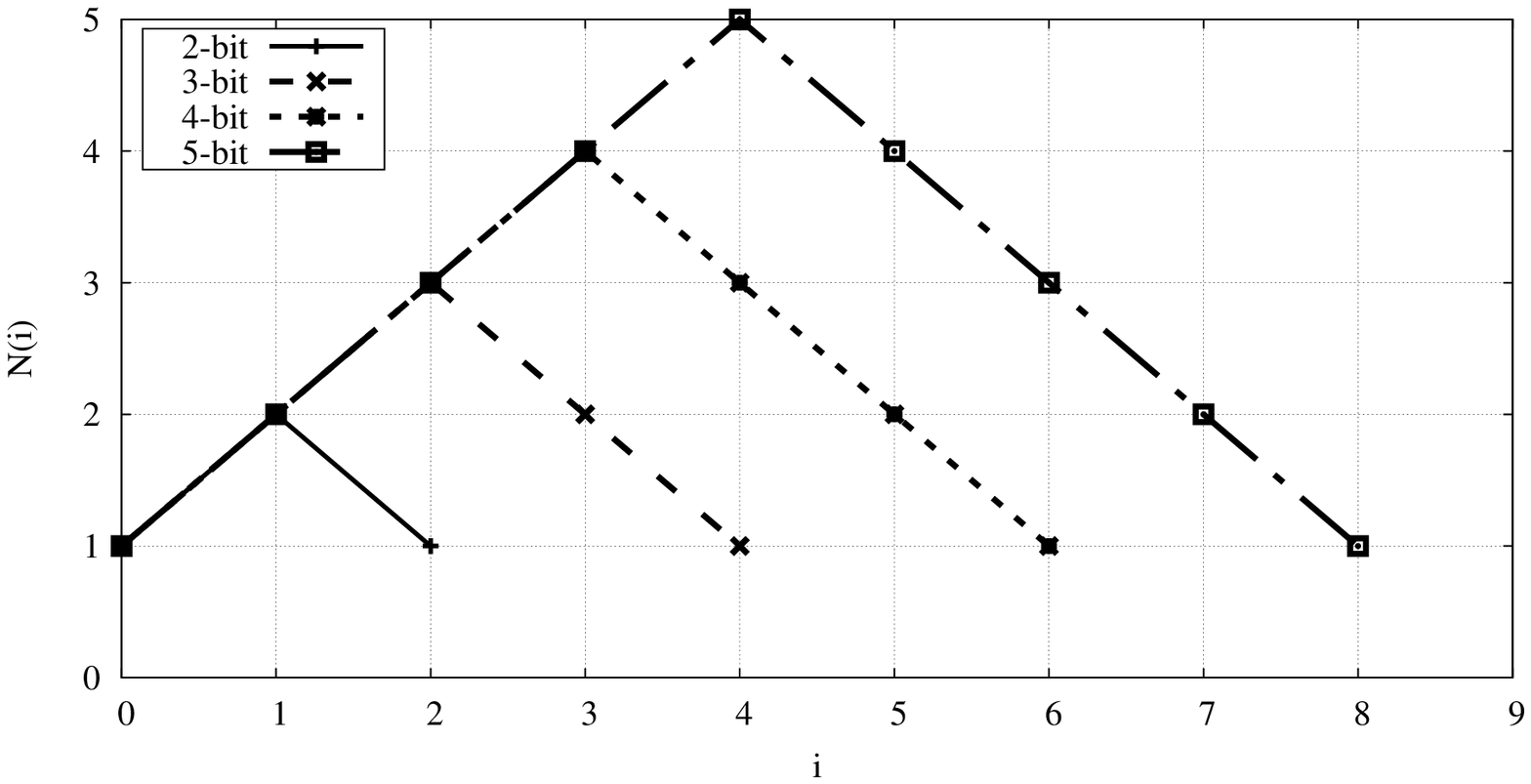}
        \caption{$F= A \cdot B$.}        
    \end{subfigure}
    \quad
    \begin{subfigure}[t]{0.23\textwidth}
        \centering
        \includegraphics[width=\textwidth]{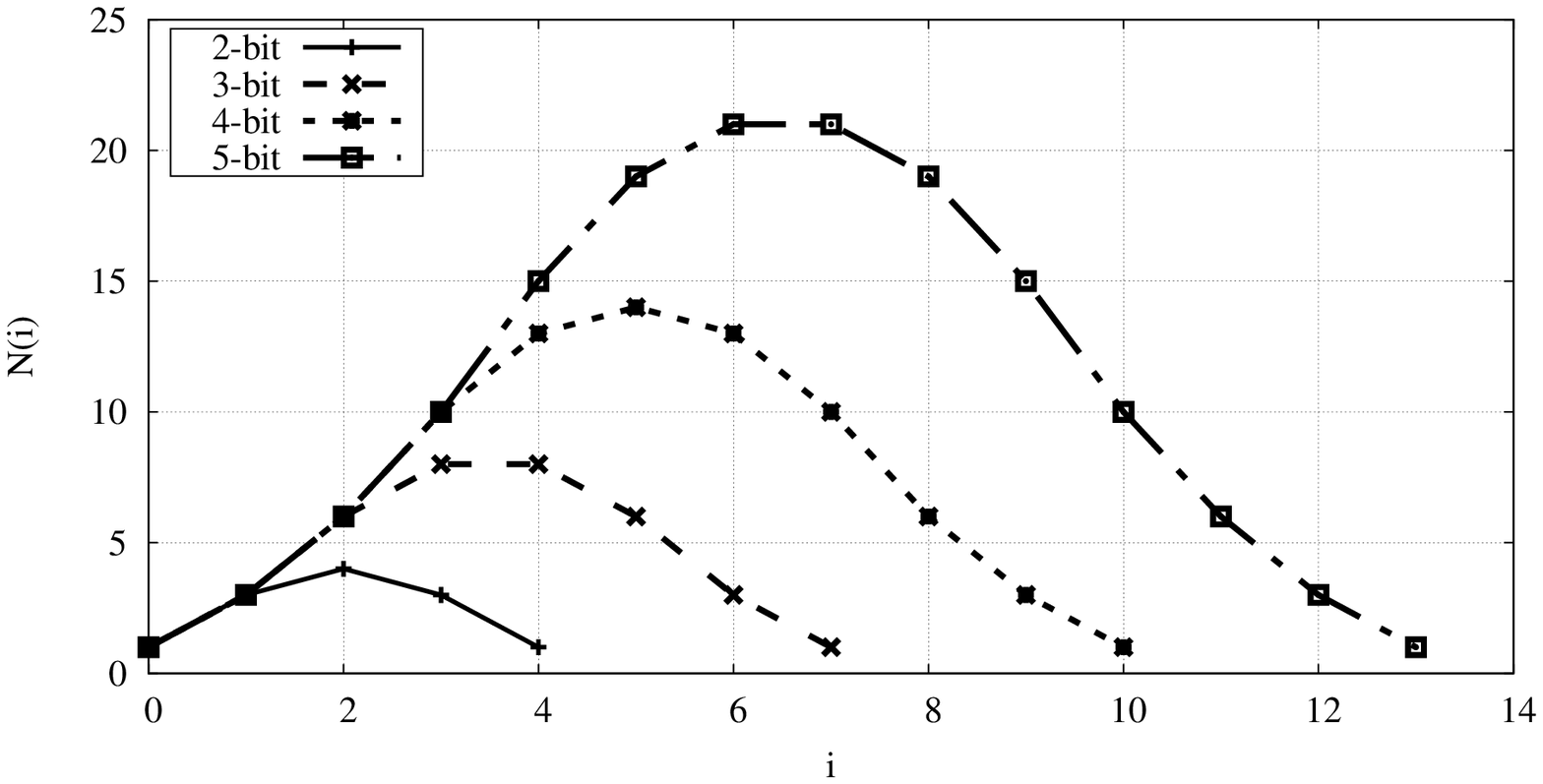}
        \caption{$F= A \cdot B \cdot C$.}
    \end{subfigure}
    \caption{Spectral diagrams for multipliers for \{ 2,3,4,5 \} bit-widths.}
    \label{fig:mult-spectra}\label{fig:Mult2op}
    \vspace{-2mm}
\end{figure}
\begin{figure}[ht]
        \centering
        \vspace{-2mm}
       \includegraphics[width=0.3\textwidth]{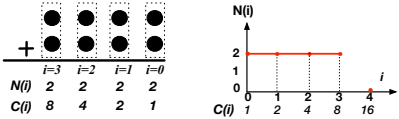}
        \caption{Spectrum of a four-bit Adder: $F= A + B$.}   
        \label{fig:Add-4bit}
        \vspace{-2mm}
    \end{figure}

Spectrum $S$ can be visualized by a graph, as shown in Figures \ref{fig:4bitMult}, \ref{fig:mult-spectra}, and \ref{fig:Add-4bit}.
The shape of the spectrum (triangle for two-input multiplier, bell curve for 3-operand multipliers, or constant line for adders, etc.) remains the same for a given arithmetic function and does not depend on the number of bits. 
Furthermore, it does not depend on the internal structure of the circuit but only on the arithmetic function it implements. A correct shape of the spectrum is one evidence of circuit correctness, but one still needs to perform canonical rewriting for final confirmation. However, an incorrect spectrum can effectively prove that the circuit is buggy (Section \ref{sec:without-rewrite}).
Figure \ref{fig:mult-spectra} shows the spectra for two-operand (2-variable spectrum) and three-operand multipliers (3-variable spectrum) for different bit-widths. 
  
Algebraic spectrum for an adder can be similarly derived.  
Clearly, for an $n$-bit binary adder with two inputs $A,B$, 
the sum $S = \sum_{i=0}^{n-1} 2^i a_i + \sum _{i=0} ^{n-1} 2^i b_i = \sum_{i=0}^{n-1} 2^i (a_i + b_i)$. Hence the number of coefficients $C_i$ with value $2^i$ is exactly two, and the spectrum is a constant function, $N_i$=2, where $i=0,...,n-1$. Again, the $(n+1)^{st}$ element of $N$ associated with the carry out bit is not shown since $N_{2n}$=0.
Algebraic spectrum for a 4-bit adder is shown in Figure \ref{fig:Add-4bit}. 
Similar formulas and graphs can be derived for other datapath operators, such as MAC, fused multiply/add operation, and others\footnote{More algebraic spectrum are available in our online spectrum gallery. \url{https://ycunxi.github.io/cunxiyu/spectrum_gallery.html}}.

\begin{figure}[hb]    
        \centering
        \includegraphics[width=0.36\textwidth]{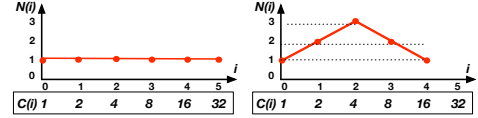}
    \caption{Spectrum of a 3-bit MAC composed of a single-variable and two-variable spectra, $S$ = \{$S_1, S_2$\}.}
    \label{fig:mac-spectrum}
\end{figure}

Note that in a monolithic arithmetic function (i.e., function composed of only one arithmetic operator) each monomial contains the same number of variables. %
For example: an adder $\sum 2^i (a_i+b_i)$ will contain only single-variable terms, regardless of the number of operands; a 2-input multiplier $\sum 2^{j+k} (a_j b_k)$ contains only two-variable terms; a 3-operand multiplier will contain only three-variable terms; etc.
However, a fused multiplier $A + B\cdot C = \sum 2^i a_i + \sum 2^{j+k} (b_j c_k)$ will contain both a single-variable terms \{$a_i$\} and two-variable terms \{$b_j c_k$\}. 
In this case the polynomial $P$ representing the function implemented by the circuit
is composed of a set of polynomials $\{P(k)\}$, where $k$ is the number of variables in each product term $p_i$. The spectrum is then computed for each value of $k$, denoted $S_k$. 
An example of such a spectrum is shown in Figure \ref{fig:mac-spectrum} for a fused multiply-add function, $A+B\cdot C$, composed of spectra $S_1$ (with single-variable monomials) and $S_2$ (with two-variable monomials).

The idea of partitioning the spectrum into components {\{$S_k$\}}, each for a different monomial size (number of variables), also applies to intermediate polynomials generated during rewriting. It can prove useful in determining when a particular arithmetic function appears in the implementation, as explained in the next section.
For example, during rewriting of a sub-expression $P=2C+S$ of a half-adder, with carry $C=a\cdot b$ and sum $S=a+b-2ab$, the expression $P=2C +a+b-2ab$ may temporarily exist  before $C$ is substituted with $ab$, which subsequently reduces $P$ to $a+b$. This means that some intermediate polynomials may map into one-variable and two-variable spectra, $S_1$, $S_2$. 
The same is true for the multiplier whose intermediate polynomials may contain monomials with three or more variables, while the final spectrum is only of $S_2$ type. 


%% file: function-extraction.tex
\subsection{Using Spectrum for Function Extraction}
\begin{figure}[hbt]    
    \centering
    \begin{subfigure}[t]{0.47\textwidth}
        \centering
        \includegraphics[width=\textwidth]{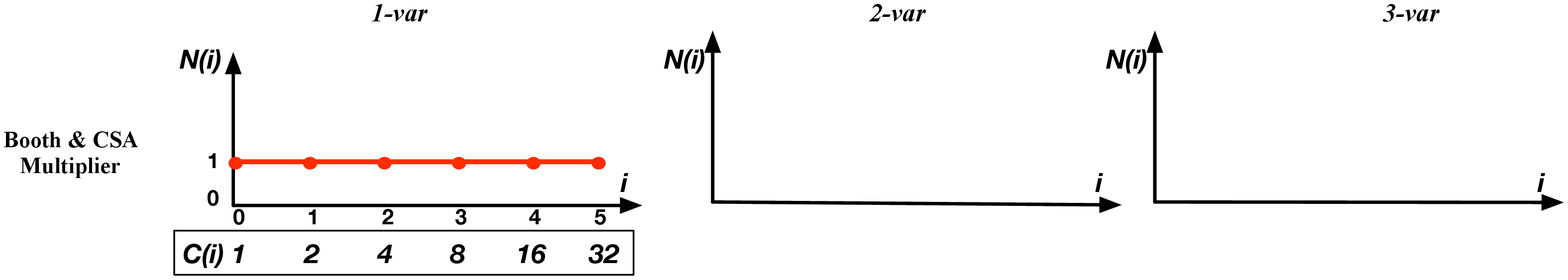}
        \caption{Initial spectrum.}        
    \end{subfigure}
    \quad
    \begin{subfigure}[t]{0.45\textwidth}
        \centering
        \includegraphics[width=\textwidth]{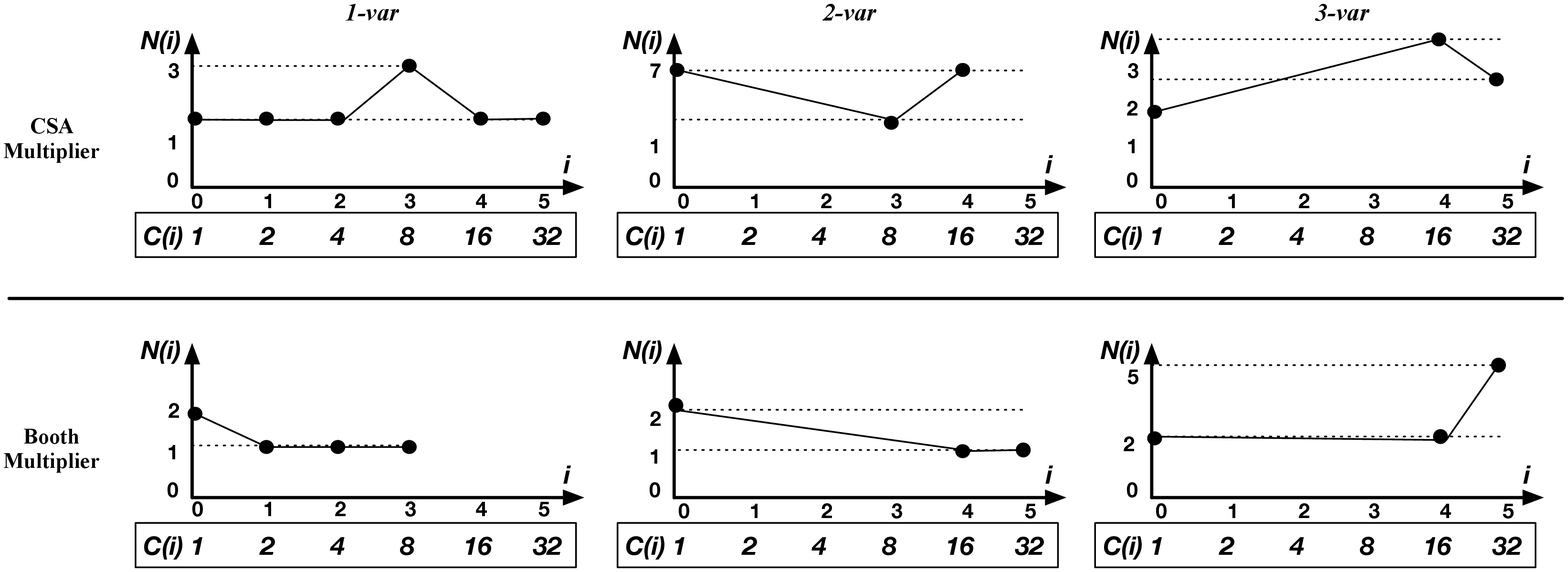}
        \caption{Intermediate spectra.}
    \end{subfigure}
    
    \quad
    \begin{subfigure}[t]{0.45\textwidth}
        \centering
        \includegraphics[width=\textwidth]{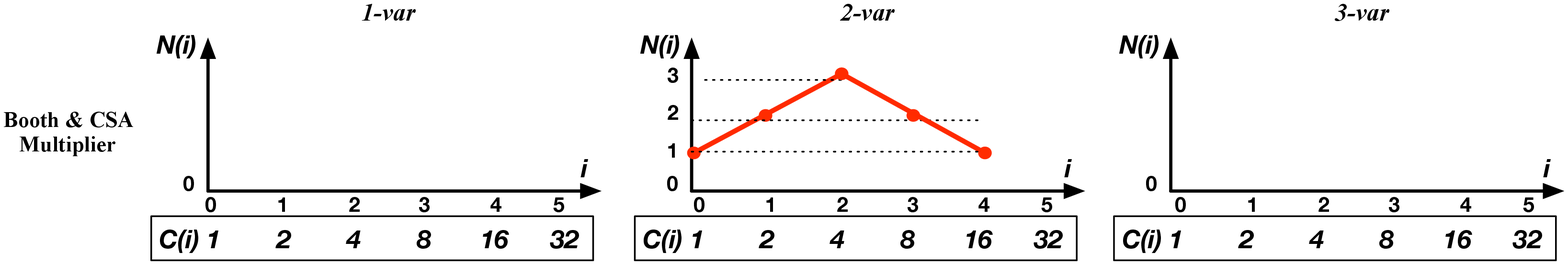}
        \caption{Final recorded spectrum.}
    \end{subfigure}
    \caption{Spectra of a three-bit Booth-multiplier and a CSA-multiplier of the four recorded expressions.}\label{fig:booth-csa-spectrum}
    \label{fig:Booth-CSA-spectrum}
    \vspace{-3mm}
\end{figure}

As mentioned earlier, the spectrum of an arithmetic circuit depends only on the arithmetic function it computes and not on its gate-level implementation. This is illustrated with an example of a 3-bit unsigned Booth and a CSA multiplier. Figure \ref{fig:Booth-CSA-spectrum} summarizes the rewriting process by showing the initial spectrum (identical for both multipliers); one intermediate spectrum for each multiplier "half way" through the rewriting process; and the final, identical spectra. 

At each step, the intermediate polynomial $P$ is divided into several sets, depending on the number of variables in its monomials, and mapped onto the corresponding spectrum $S_k$. The first column in the figure represents $S_1$, the second represents $S_2$, and the third represents $S_3$. The initial polynomial $P=Sig_{out}$, contains only single-variable monomials, namely $z_{0}$+$2z_1$+$4z_2$+$8z_3$+$16z_4$+$32z_5$, corresponding to the word-level encoding of the output, and is the same for both multipliers. Hence the initial spectrum is the same for both, as shown in Figure \ref{fig:Booth-CSA-spectrum}(a).
During rewriting, the size of some monomials increases to 2 or 3 variables, which is captured by the spectra $S_2$ and $S_3$, shown in Figure \ref{fig:Booth-CSA-spectrum}(b). Upon the completion of the rewriting the polynomial associated with the primary inputs, $P=Sig_{in}$, contains only monomials of size 2, in both multipliers. Hence, the spectrum $S_2$ of the two multipliers is identical, if the circuit is a bug-free multiplier.
 As expected, the {\it intermediate} spectra for the two multipliers are different, since they are implemented using different algorithms and have different internal structures. However, the final spectra of both circuits at the primary inputs match the spectrum $S_2$ of the multiplication, showing that they both implement the multiplication function. A buggy circuit may contain monomials with a larger number of variables with coefficients that do not match those of the correct circuit, which will be an indication of a bug.
 %
%
\subsection{Using Spectrum in Arithmetic Verification} \label{sec:uniqueness}
According to Definition 1, \textit{algebraic spectrum} is a more abstract and compact representation of an arithmetic function compared to a polynomial representation. However, the spectrum alone, as defined here, is {\it not canonical}. This is because it only deals with the distribution of coefficients and does not differentiate between the variables in the product terms $p_i$. As a result, different polynomials may map into the same spectrum, as shown in this example.

{\bf Example 2}: Let $P_1$ and $P_2$ be the polynomial expressions of two multiplications, $P_1=(a_0+2a_1)(b_0+2b_1) = a_0b_0 +2a_0b_1 +2a_1b_0 +4a_1b_1$, and $P_2=(a_1+2a_0)(b_0+2b_1) = a_1b_0 +2a_1b_1 +2a_0b_0 +4a_0b_1$; obviously they are \textit{not functionally equivalent}. The difference between $P_1$ and $P_2$ is in the bit composition of the first operands. Yet, the spectrum of both polynomials are identical, $S = S_2 = \{ (1,1), (2,2), (1,4) \}$, each with distinct coefficients $C= \{1, 2, 4\}$. Hence, such defined spectrum is not canonical. 

To make the representation canonical and useful for verification, we need to relate it to the input variables, while avoiding computing the input signature by the expensive backward rewriting of the entire circuit. This can be accomplished by local rewriting of the polynomial associated with the spectrum, as explained next.  


%% file: adder-tree-extract.tex
\section{Spectrum Computation without Explicit Rewriting}\label{sec:without-rewrite}

In this section we introduce a method that extracts algebraic spectrum without performing explicit rewriting. We shall rely here on a {\it functional representation} of the circuit using an And-Inverter Graph (AIG) representation of the gate-level circuit. 
In particular we will use AIG to propagate the weights through adder trees, present in some form in most arithmetic circuits.

\subsection{Adder-tree Extraction {and Coefficient Propagation}}

AIG provides a compact way to represent combinational logic circuits. It is a directed acyclic graph whose internal nodes represent two-input AND functions and the edges are labeled to indicate an optional signal inversion \cite{mishchenko:2006-dag}\cite{kuehlmann1997equivalence} \cite{abc-link}. 
Any Boolean network can be transformed into an AIG using DeMorgan's law. 
We will use AIG structure to extract adder trees by detecting $XOR3$ and $MAJ3$ functions with identical inputs since they represent the {\it sum} and the {\it carry} of the adder, respectively. ABC provides a method to extract adder-tree structure from a gate-level netlist\cite{abc-link}. It does it by computing \textit{cuts}, sets of AIG nodes called \textit{leaves}, such that each path from PIs to $n$ passes through the leaf nodes. A cut is $K$-feasible if the number of leaves does not exceed $K$. This approach, implemented by an ABC procedure \textit{\&atree}, proceeds as follows:
\begin{itemize}
\item Compute 3-feasible cuts of AIG nodes and their truth tables.
\item Store the cuts in the hash table ordered by their inputs.
\item Detect pairs of 3-input cuts with identical inputs, such that the Boolean functions of the two cuts with shared inputs belong to the \textit{NPN} classes of XOR3 and MAJ3 \cite{HuangWNM13}.
\end{itemize}
As soon as the XOR3 and MAJ3 pairs are detected, the HAs and FAs are automatically extracted.
Details are provided in \cite{HuangWNM13}.

\begin{figure}[!htb]
  \centering
    \includegraphics[width=0.33\textwidth]{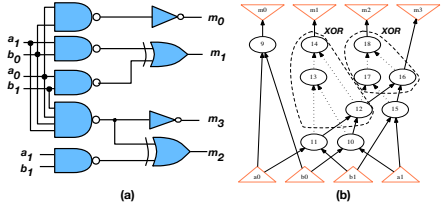} 
    \caption{A synthesized two-bit multiplier. (a) gate-level netlist; (b) AIG representation. Values inside the nodes represents node names.}
    \vspace{-2mm}
    \label{fig:adder-tree}
\end{figure}

\begin{figure}[!htb]
  \centering
    \includegraphics[width=0.37\textwidth]{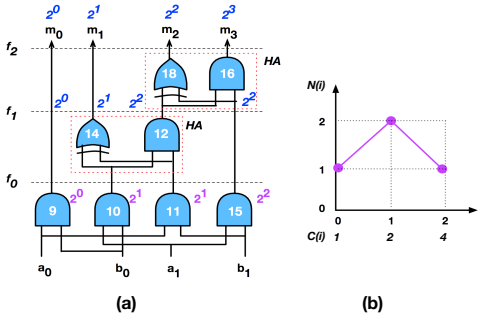} 
    \caption{Coefficient propagation in a 2-bit multiplier: (a) Netlist with adder-tree detected; (b) Constructed algebraic spectrum $S=S_2$.}
    \label{fig:adder-tree-spectrum}
    \vspace{-2mm}
\end{figure}

Our approach to compute spectrum by extracting adder-tree is based on the observation that arithmetic circuits, such as multipliers, are implemented with an adder-tree and a partial product generator, in some form. Extraction of adder trees has important advantage over the computation of individual gates since the adder function can be represented by a linear relation:
$a+b+c_{in} = 2C + S$,
where $a, b, c_{in}$ are the binary inputs and $C,S$ are the carry-out and sum of of the full adder (FA), respectively. Similar formula can be obtained for a half-adder (HA), with $c_{in}=0$.
With this, the signal {\it weights}, represented by coefficients $C_i$, needed to construct the spectrum can be computed by simply propagating the weights from the known linear polynomial of the output signature $Sig_{out} = \sum 2^i r^i$ through the adder tree, until they reach the non-linear partial product generator logic. 
%
During the backward propagation, the weight of the \textit{carry} bit of HA/FA is always 2$\times$ the weights of the inputs (which always have the same weight), and the weight of the \textit{sum} bit is the same as the weight of the inputs.
Once the propagation reaches partial products, standard backward rewriting is applied, but now to a relatively shallow logic. This can significant reduce the computation efforts compared to backward rewriting on adder-tree \cite{yu2018fast}, since weight propagation requires much less computations than regular backward rewriting. 
Propagation of the weights in a Booth multiplier, which contains recorded partial products, is also possible; it is discussed later in Example 4.
We first illustrate the algorithm of constructing spectrum with an example of a 2-bit multiplier, Figures \ref{fig:adder-tree} and \ref{fig:adder-tree-spectrum}.

\textbf{Example 3 (CSA multiplier)}: A mapped gate-level netlist of a 2-bit CSA-multiplier and its AIG are shown in Figure \ref{fig:adder-tree}. Here $n_i$ denotes node labeled $i$ in the figure. Computing 3-feasible cuts in the AIG reveals the following matching:
node $n_{14}$ is an \textit{XOR3} and node $n_{12}$ is a \textit{MAJ3} on shared inputs (\textit{$n_{10}$, $n_{11}$, 0}). 
Similarly, nodes $n_{18}$ and $n_{16}$ form an \textit{XOR3, MAJ3} pair on inputs (\textit{$n_{12}$, $n_{15}$, 0}). 
This corresponds to two half-adders (HA), composed of gates (18, 16), and gates(14, 12), shown in Figure \ref{fig:adder-tree-spectrum}(a).
The weights of all the signals are then propagated backward from PO to PI in reverse-topological order, using linear expression $2C+S = a+b$ for the HAs.

First, the weights (signal coefficients) of HA(18,16) are propagated to cut $f_1$. 
As a result, the weight of gate 18 (signal $S$) is $2^2$. Hence, both inputs of gates 18,16 must have weight $2^2$. Similarly, at cut $f_0$, the weights of inputs of gates 14 and 12 are $2^1$. The algorithm terminates at this point since there are no more HA or FA nodes. The spectrum, shown in Figure \ref{fig:adder-tree-spectrum}(b) represents the distribution of coefficients at cut $f_0$, with outputs of gates $9,10,11,15$. The spectrum indicates that the circuit is a 2-bit multiplier, but, as noted earlier, 
we need additional steps to find the composition of the operands to confirm the results. On the other hand, the incorrect spectrum can be used to quickly determine that the circuit is buggy, i.e., it does not satisfy the expected arithmetic function. This is explained by the following theorem regarding the necessary condition for a circuit to be a multipliers.


\textbf{Theorem:} 
The circuit is a multiplier only if its spectrum $S$ is a single 2-variable spectrum $S_2$ that satisfies Eq.(\ref{eq:mult-spectrum-formula}). 

\textbf{Proof:} Assume that $S$ contain other spectra $S_i$ than $S_2$, i.e., $i = 1$, or $i > 2$. Then, according to Definition 1, the functional specification $F$ of the circuit must include at least one monomial with a single variable or with more than two variables, which contradicts the definition of multiplication (Eq.\ref{eq:spec-multiplication}).
Similarly, if $S = S_2$, but $S_2$ does not match Eq.(\ref{eq:mult-spectrum-formula}) of the multiplier's spectrum, then some of the coefficients do not match the definition of the multiplication operation  (Eq.\ref{eq:spec-multiplication}), and hence it cannot be a correct multiplier.


\subsection{Extracting Arithmetic Function from the Spectrum}
In order to get the full information and extract the true arithmetic function of the circuit, a {\it canonical polynomial} expression in terms of PI needs to be derived. 
This can be readily accomplished by combining the computation of the spectrum with local rewriting of the associated polynomial, as explained by the following.

{\bf Definition 2:}
Let $S$=\{ $(N_1,C_1), \dots , (N_m,C_m)$ \} be an algebraic spectrum with coefficients $C=\{C_1,...,C_m \}$. By definition, each element ($N_i,C_i$) of $S$ is associated with $N_i$ monomials, \{$p_{i}^1,p_{i}^2,...,p_{i}^{N_i}$\}, each with a coefficient $C_i$. The polynomial corresponding to spectrum $S$, with variables representing the monomials $p_i$, is called a {\it Spectral Polynomial}, $SP(S)$, and has the following form: 
$SP(S) = \sum_{i=1}^{m} (\sum_{j=1}^{N_i} C_i p_i^{j} )$.

By construction, it is a linear polynomial reconstructed from the spectrum that represents a polynomial expression of a cut at a set of variables $\{p_i\}$. To obtain the input signature, $Sig_{in}$ we just need to express each variable $p_i$ in terms of the primary inputs PI, which can be done by backward rewriting.
In the case of an adder, each $p_i$ is already a primary input, PI, so the $SP(S)$ is the input signature, $Sig_{in}$.
For a standard, non-Booth multiplier, with simple partial products, each variable $p_i$ is a product of some input variables $a_jb_k$. And in the case of a Booth multiplier, each $p_i$ can be expressed as a non-linear polynomial in terms of PI, typically a sum of products of the input variable (see Example 5).

{\bf Example 4}:
Consider again the 2-bit multiplier and its spectrum in Figure 7. The spectrum derived by the adder-tree extraction corresponds to the cut $f_0$ and has the following form: $\{(1,1), (2,2), (1,4)\}$. The corresponding spectral polynomial is $SP = p_1 + 2p_2 + 2p_3 +4p_4$, where the individual variables $p_i$ correspond to outputs of gates $9,10,11,15$, respectively. They can be traced by backward rewriting to PI as follows: 
$p_1=a_0b_0; p_2=a_0b_1; p_3=a_1b_0; p_4=a_1b_1$. This results in the input polynomial
$Sig_{in} = a_0b_0 + 2a_0b_1 + 2a_1b_0 + 4a_1b_1 $, a canonical representation of the 2-bit multiplier circuit.

In summary, the idea is to first generate the spectrum of the linear portion of the circuit and then use it to derive the input signature by polynomial rewriting based on Definition 2. In contrast to the original rewriting approach, the polynomial rewriting is done here only on a local non-linear portion of the circuit. By combining spectral analysis and local backward rewriting we can generate a canonical arithmetic function representation, $Sig_{in}$, and use it to solve the verification and abstraction problems.
%
%

\subsection{Handling Booth Multipliers}

We conclude this section by analyzing the application of our approach to Booth multipliers.
The logic of partial product generators depends on the multiplication algorithm used in constructing the multiplier. For example, CSA-multiplier uses an AND array, while Booth-multiplier uses recoded partial products. Nonetheless, once the adder-tree is detected, the algebraic spectrum is extracted in the same fashion, regardless of the type of the multiplier. Booth-encoded multiplier has more complex partial product logic with fewer product terms in order to minimize area and the delay of the multiplier. 
The following example illustrates our approach of spectrum construction and polynomial generation by fast local rewriting using a 3-bit Booth-multiplier.

\textbf{Example 5 (3-bit radix-4 Booth-multiplier):}  Polynomial expressions of all the partial products of this Booth multiplier are shown in Eq.(\ref{eqn:booth}). Arithmetic function of the circuit is the weighted sum of these partial products, with the weights shown on the left.
Note that some of the partial products contain three variables. However, it can be shown that those products are {\it redundant}, because they cancel each other in the weighted sum. In Eq.(\ref{eqn:booth} the underlined 3-variable terms will be cancelled.
For example, $a_{2}b_{1}b_{2}$, which appears in $pp_{31}$ and $pp_{21}$, will get cancelled in the partial sum $2^5 pp_{31} + 2^4 pp_{21}$, so that  $2^4a_{2}b_{1}b_{2}$+$2^3(-2a_{2}b_{1}b_{2})$=0. The same is true for other 3-variable terms, resulting in a 2-variable spectrum only. The remaining 2-variable terms form the final polynomial: $a_ob_0+ 2a_0b_1+ 2a_1b_0+ 4a_1b_1+ 4a_2b_0 +4a_0b_2 +8a_2b_1 +8a_1b_2 +16a_2b_2$, representing a 3-bit multiplication. This polynomial will be derived from the spectrum polynomial, as discussed in Example 4, 

\vspace{-1mm}
{
\scriptsize
\begin{eqnarray}\label{eqn:booth}
\begin{aligned}
\scriptsize
& 2^2 \cdot 2^3: pp_{31} =  \underline{a_{2}b_{1}b_{2}}\\
& 2^2 \cdot 2^2: pp_{21} = \underline{-2a_{2}b_{1}b_{2}}+\underline{a_{1}b_{1}b_{2}}+a_{2}b_{1}+a_{2}b_{2} \\
& 2^2 \cdot 2^1: pp_{11} = \underline{-2a_{1}b_{1}b_{2}}+\underline{a_{0}b_{1}b_{2}}+a_{1}b_{1}+a_{1}b_{2}\\
& 2^2 \cdot 2^0: pp_{01} = \underline{-2a_{0}b_{1}b_{2}}+a_{0}b_{1}+a_{0}b_{2}\\
& 2^3: pp_{30} = \underline{a_{2}b_{0}b_{1}}-a_{2}b_{1}\\
& 2^2: pp_{20} = \underline{-2a_{2}b_{0}b_{1}}+\underline{a_{1}b_{0}b_{1}}-a_{1}b_{1}+a_{2}b_{0}\\
& 2^1: pp_{10} = \underline{-2a_{1}b_{0}b_{1}}+\underline{a_{0}b_{0}b_{1}}-a_{0}b_{1}+a_{1}b_{0}\\
& 2^0: pp_{00} = \underline{-2a_{0}b_{0}b_{1}}+a_{0}b_{0}
\end{aligned}
\end{eqnarray}
}

%

%% file: results.tex
\vspace{-2mm}
\section{Results}

\input{pre-post-table.tex}


The technique described in this paper has been implemented in C++ and integrated with the ABC tool \cite{abc-link}. The program takes as input the gate-level netlist in Verilog, BLIF or AIG format, and produces algebraic spectrum and the final polynomial, $Sig_{in}$, of the circuit. The experiments involved computing the spectrum for various multipliers and arithmetic combinational datapath circuits in the original (non-optimized) and synthesized versions, with synthesis performed by ABC. The benchmarks involve the CSA and radix-4 Booth multipliers, taken from \cite{yu:2016-tcad-verification}\cite{homma2006formal}\cite{ritirc2017column}. The experiments were conducted on a PC with Intel(R) Xeon CPU E5-2420 v2 2.20 GHz x12 with 32 GB memory. 

Two types of experiments were performed: 1) {\bf verification}, in which the computed polynomial $Sig_{in}$ is compared with the given specification polynomial; and 2) {\bf function abstraction}, where the computed spectrum is analyzed to determine the type of arithmetic function implemented by the circuit. The {\it verification} results are compared with the state-of-the-art approaches presented in \cite{yu:2016-tcad-verification}\cite{ritirc2017column}\cite{ritirc2018improving}. 
For {\it word-level abstraction}, our approach is compared with the simulation graph-based technique \cite{soeken:2015simulationgraph} and computer algebra method of \cite{yu:2016-abstraction}. 
The comparison with the contemporary formal methods such as SAT, SMT and commercial tools are not provided in this paper; computer algebraic approach has already been shown to be  orders of magnitude faster than those techniques \cite{yu:2016-tcad-verification}. 


\input{dac2018-mult-result.tex}

{\bf Verification results} for the original and synthesized multipliers are shown in Tables \ref{tbl:pre-post-tbl} and \ref{tbl:many_mult}. The CPU times are compared to \cite{yu:2016-tcad-verification}\cite{sayedformal:date-2016}\cite{ritirc2017column}\cite{ritirc2018improving}. 
Multipliers \textit{btor} are generated from Boolector \cite{niemetz:2015boolector}; CSA-multipliers are taken from \cite{yu:2016-tcad-verification}. 
The multipliers in the third and fourth rows of Table \ref{tbl:many_mult} are AOKI multipliers \cite{homma2006formal}, used in works of \cite{sayedformal:date-2016}\cite{ritirc2017column}\cite{ritirc2018improving}. The naming of AOKI multipliers is explained in \cite{sayedformal:date-2016}. Multipliers \textit{abc} and \textit{abc-booth} are generated by ABC, using command \textit{[gen -N -m]} and \textit{[\%blast -b]}.
The results show that the verification based on spectral method is significantly faster than the other methods.
Furthermore, while it has been previously shown that synthesis can adversely affect the verification efficiency \cite{yu:2016-tcad-verification}\cite{ritirc2017column}, the spectral method is equally efficient for both synthesized and non-synthesized multipliers. 
However, three failure cases were observed while applying spectral method to AOKI multiplier circuits. They include circuits \textit{bp-wt-cl} (Booth multiplier with Wallace-tree and Carry-look-ahead adder), \textit{sp-ar-rc-dc2}; and \textit{bp-ar-rc-dc2} (optimized Booth and standard multipliers with Ripple Carry Adder). For these circuits, the process of constructing spectra did not work due to the presence of {\it unstructured adder tree} (UAT) component that could not be handled by the ABC adder extraction feature. Note that verifying large Booth-multiplier is much faster than verifying the CSA and ABC-generated multipliers. This is because Booth multiplier has much smaller adder-tree and significantly fewer gates. Specifically, the tested 1024-bit CSA-multiplier has over \underline{12 million} gates, and 1024-bit Booth-multiplier has only 3million gates.

We also tested the application of our spectral method on {\it buggy} multipliers. The last row of Table \ref{tbl:many_mult} includes two 256-bit buggy multipliers, \textit{abc-buggy} and \textit{abc-booth-buggy}. The bugs are introduced randomly inside the adders of these two multipliers. As a result, the clean adder-tree could not be detected, because it does not exist. One interesting observation is that in the AIG of a buggy circuit, the place where the adder-tree breaks is close to the bug location. This can be used in the future to identify and find bug location.

{\bf Abstraction results}: extracting word-level specifications from gate-level complex arithmetic circuits are shown in Table \ref{tbl:complex-circuits}. We use three types of circuits that are constructed with multiplication and addition, and a three-operand multiplier. The multiplications in these datapaths are implemented using ABC-generated multipliers. It shows that our approach can efficiently identify the word-level operations in the gate-level datapaths. In contrast, the approach of \cite{soeken:2015simulationgraph} cannot tell whether there exists multiplication or addition in these circuits; and our approach is much faster than \cite{yu:2016-abstraction}. 

\input{complex-circuits.tex}

%% file: pre-post-table.tex
\begin{table}[!htb]
\scriptsize
\centering
\caption{CPU runtime (seconds) of verifying pre- and post-synthesized {\bf gate-level} CSA multipliers compared to techniques in \cite{yu:2016-tcad-verification}\cite{sayedformal:date-2016}\cite{ritirc2017column}\cite{ritirc2018improving}; source: gate-level netlist from \cite{yu:2016-tcad-verification}. \textit{MO} = Memory out of 16 GB. \textit{TO} = Time Out (3 hrs). ES = Error state reported.}
\vspace{-1mm}
\label{tbl:pre-post-tbl}
\begin{tabular}{|r|r|c|c|c|r|c|c|c|r|}
\hline
\multicolumn{1}{|c|}{\multirow{2}{*}{Size}} & \multicolumn{5}{c|}{Pre-synthesized} & \multicolumn{4}{c|}{Post-synthesized} \\ \cline{2-10} 
\multicolumn{1}{|c|}{} & \multicolumn{1}{c|}{\cite{yu:2016-tcad-verification}} & \cite{sayedformal:date-2016} &\cite{ritirc2017column}&\cite{ritirc2018improving}& \multicolumn{1}{c|}{\textbf{Ours}} & \cite{yu:2016-tcad-verification} &\cite{ritirc2017column}&\cite{ritirc2018improving}& \multicolumn{1}{c|}{\textbf{Ours}} \\ \hline
64 & 1.9 & TO & \multicolumn{1}{r|}{801} & \multicolumn{1}{r|}{4.0} & 0.1 & \multicolumn{1}{r|}{5.5} & \multicolumn{1}{r|}{1073} & \multicolumn{1}{r|}{418} & 0.1 \\ \hline
128 & 8.1 & - & ES & ES & 0.8 & \multicolumn{1}{r|}{40} & ES & ES & 0.9 \\ \hline
256 & 33 & - & - & - & 7.8 & \multicolumn{1}{r|}{285} & - & - & 8.4 \\ \hline
512 & 130 & - & - & - & 30 & MO & - & - & 42 \\ \hline
1024 & \multicolumn{1}{c|}{MO} & - & - & - & 9638 & MO & - & - & 9817 \\ \hline
\end{tabular}
\end{table}

%% file: dac2018-mult-result.tex
\begin{table}[!htb]
\centering
\scriptsize
\caption{Runtime (seconds) of verifying multipliers implemented using different architectures; source: \textbf{AIG} from \cite{yu:2016-tcad-verification}\cite{sayedformal:date-2016}\cite{ritirc2017column}\cite{ritirc2018improving}. \textit{MO} = Memory out of 16 GB. \textit{TO} = Time Out (3 hrs). \textit{UAT} = Unstructured adder-tree detected. ES = Error state reported.}
\label{tbl:many_mult}
\begin{tabular}{|c|l|c|c|c|c|r|}
\hline
n-bit & MULT benchmarks & {\cite{yu:2016-tcad-verification}} & \multicolumn{1}{l|}{\cite{sayedformal:date-2016}} & \multicolumn{1}{l|}{\cite{ritirc2017column}} & \cite{ritirc2018improving} & \multicolumn{1}{c|}{\bf Ours} \\ \hline
\multirow{4}{*}{128} & \textit{\begin{tabular}[c]{@{}l@{}}btor; btor-resyn3;\\ abc; abc-resyn3; \\ CSA; ~CSA-resyn3;\end{tabular}} & MO & TO & ES & ES & 1.5 \\ \cline{2-7} 
 & \textit{\begin{tabular}[c]{@{}l@{}}abc-booth;\\ abc-booth-resyn3\end{tabular}} & MO & TO & ES & ES & 0.5 \\ \cline{2-7} 
 & \textit{\begin{tabular}[c]{@{}l@{}}sp-ar-rc [AOKI]\end{tabular}} & - & TO & ES & ES & 1.5 \\ \cline{2-7} 
 & \textit{\begin{tabular}[c]{@{}l@{}}bp-ar-rc-dc2(resyn3) [AOKI];\\ sp-ar-rc-dc2(resyn3) [AOKI]\end{tabular}} & - & - & - & - & UAT \\ \hline
\multirow{3}{*}{256} & \textit{abc; ~abc-resyn3} & MO & TO & - & - & 14 \\ \cline{2-7} 
 & \textit{abc-booth; ~abc-booth-resyn3} & MO & TO & - & - & 3.5 \\ \cline{2-7} 
 & \textit{abc-buggy; ~abc-booth-buggy} & - & - & - & - & UAT \\ \hline
\multirow{2}{*}{1024} & \textit{abc; ~abc-resyn3} & - & - & - & - & 9482 \\ \cline{2-7} 
 & \textit{\begin{tabular}[c]{@{}l@{}}abc-booth; abc-booth-resyn3\end{tabular}} & - & - & - &  & 139 \\ \hline
\end{tabular}
\end{table}

%% file: complex-circuits.tex
\begin{table}[!htb]
\centering
\scriptsize
\vspace{-2mm}
\caption{Results of extracting word-level specification from complex arithmetic circuits. \textit{TO} = TIME OUT (3600 s); \textit{error} = Wrongly reported that no multiplication nor addition component exist; \textit{TO*}: finished in 23,760 s.}
\vspace{-1mm}
\label{tbl:complex-circuits}
\begin{tabular}{|l|l|l|l|l|}
\hline
\multicolumn{1}{|c|}{256-bit} & \multicolumn{1}{c|}{\cite{soeken:2015simulationgraph}} & \multicolumn{1}{c|}{\cite{yu:2016-abstraction}} & \multicolumn{2}{c|}{\bf Ours} \\ \hline
\textit{F=A$\times$B+C} & error & TO* & 1$\times$mult;1$\times$add & 44.7 s \\ \hline
\textit{F=A$\times$(B+C)} & error & TO & 2$\times$mult & 45.1 s \\ \hline
\textit{F=A$\times$B$\times$C} & error & TO & 1$\times$mult3 & 68.5 s \\ \hline
\end{tabular}
\end{table}

%% file: conclusions.tex
\vspace{-4mm}
\section{Conclusions}

The paper presents a novel {\it spectral} analysis method for arithmetic circuit verification. Our approach extracts and analyzes an arithmetic function implemented by the circuit by efficient computation of the input signature polynomial; explicit algebraic rewriting is largely avoided by propagating signal weights through an adder tree using AIG adder-tree extraction. The method described here can be used for word-level function extraction of an arithmetic circuit and for functional checking of the gate-level circuit against its polynomial specification. The experimental results show that it outperforms the currently known approaches in verification and abstraction for gate-level arithmetic circuits.

This work is naturally limited to integer combinational arithmetic circuits whose function can be expressed by polynomials; it is not directly applicable to dividers, transcendental, and other functions that do not have a closed-form polynomial representation. 
The benefit of fast spectrum computation and adder-tree extraction strongly depends on the structure of the circuit; the more unstructured the adder-tree portion is, the more burden will fall on algebraic rewriting instead of the spectrum computation. For architectures with highly unstructured (or absent) adder trees the adder-tree extraction may even fail, and the size of intermediate polynomials that need to be computed instead may become prohibitively large. 

Applying spectral method to {\it debugging} and analysis of {\it faulty circuits} requires more insight. In principle, a bug in a circuit will manifest itself by the fact that the final input polynomial does not match the expected spectral specification. However, those circuits are even more prone to failing the adder-tree extraction and can cause exponential blowup in the polynomial size during rewriting. 
In any case, the method can be used to quickly {\it disprove} that whether the circuit implements the expected {\it type} of the function, such as multiplication. 
